# The Unintended Consequences of Stay-at-Home Policies on Work Outcomes: The Impacts of Lockdown Orders on Content Creation

Xunyi Wang[1], Reza Mousavi[2], Yili Hong[3]

[1] Baylor University, [2] University of Virginia, [3] University of Houston

## Abstract

The COVID-19 pandemic has posed an unprecedented challenge to individuals around the globe. To mitigate the spread of the virus, many states in the U.S. issued lockdown orders to urge their residents to stay at their homes, avoid get-togethers, and minimize physical interactions. While many offline workers are experiencing significant challenges performing their duties, digital technologies have provided ample tools for individuals to continue working and to maintain their productivity. Although using digital platforms to build resilience in remote work is effective, other aspects of remote work (beyond the continuation of work) should also be considered in gauging true resilience. In this study, we focus on content creators, and investigate how restrictions in individual's physical environment impact their online content creation behavior. Exploiting a natural experimental setting wherein four states issued state-wide lockdown orders on the same day whereas five states never issued a lockdown order, and using a unique dataset collected from a short video-sharing social media platform, we study the impact of lockdown orders on content creators' behaviors in terms of content volume, content novelty, and content optimism. We combined econometric methods (difference-in-differences estimations of a matched sample) with machine learning-based natural language processing to show that on average, compared to the users residing in non-lockdown states, the users residing in lockdown states create more content after the lockdown order enforcement. However, we find a decrease in the novelty level and optimism of the content generated by the latter group. Our findings have important contributions to the digital resilience literature and shed light on managers' decision-making process related to the adjustment of employees' work mode in the long run. Our findings also have important implications for designing digital platforms that are used for building resilience in an ecosystem.

# 1. Introduction

The COVID-19 pandemic has ravaged the world, infecting over 52 million individuals and claiming the lives of 1.3 million as of November 2020. As a response, governments have implemented various policies. Among the policies enforced by state authorities to curb the spread of the pandemic, lockdowns have been considered as the last resort and the most effective. These lockdowns, however, impose physical restrictions for the individuals in the jurisdictions. From missing on visiting family and friends to the inability to shop in malls, the lockdowns drastically changed the daily lives of individuals. In the presence of the lockdown restrictions, many individuals were also forced to work remotely via digital platforms. While offline workers are experiencing significant challenges, digital technologies have provided ample tools for individuals to continue working and to maintain their productivity. These digital platforms have been used to build resilience and aid individuals to quickly recover from or adjust to the major disruptions (e.g., lockdowns) caused by the pandemic.

Although it appears that the digital platforms are effective in building resilience by overcoming the restrictions related to geographical boundaries and aiding individuals to continue working on their tasks remotely (Huang et al. 2020), they may not address other unintended consequences of the disruptions on individuals' work. For instance, individuals can use the Zoom application to continue their work meetings (avoid disruptions); however, attending Zoom meetings from a new work environment (their homes) could have consequences on their work. For instance, research shows that individuals are exhausted at the end of the workday due to the rise of "Zoom fatigue" (Fosslien and Duffy 2020). Thus, the resilience created by digital platforms (such as the Zoom application) should not be solely interpreted as the continuation of work after a disruption. Rather, it should be interpreted as the extent to which digital platforms truly aid individuals to continue their work the same way they used to before the disruption. In this perspective, not only the continuation of work is important, but also being able to produce the same outputs as before is important. To understand to what extent digital platforms can build resilience during lockdowns, we first need to study the consequences of lockdowns in terms of individuals' work outcomes. The consequences for work outcomes could be related to their work productivity, work quality, and psychological state. On the one hand, during the lockdowns, individuals have more time to work at home as the lockdowns restrict their activities outside of their homes, but on the other hand, they may feel tired and weary of repeating the same activities from the same physical environment (their homes), which may affect work quality as well as their psychological state.

In this study, we examine the impact of working from home (as a consequence of lockdowns) on the digital resilience in work outcomes in the context of content-creation platforms. We specifically focus on content creators as they are an emerging group of individuals who are increasingly gaining prominence. Given the slowdown in the traditional media, entertainment, and sports industry, individual content creators have become an important workforce satisfying the demand for creative content. In particular, we strive to understand individuals' content creation behavior on a video creation and sharing platform, TikTok. TikTok is one of the fastest-growing social media platforms globally, which allows content creators to post original content in the form of short videos. According to CNBC, TikTok's user base has grown 800% from January 2018 to July 2020.[1] In this study, we are interested in a holistic approach to gauge digital resilience by comparing content creators' work outcomes before and after the lockdowns. Given that prior research suggests that resilience is a multi-faceted concept (Connor and Davidson 2003), we gauge digital resilience in work outcomes using three distinct constructs that are essential in content creation industry: 1- productivity, 2- novelty, 3- and optimism (positive emotions). Bearing this in mind, we seek to understand how state-wide lockdowns impacted the work outcomes of social media content creators in terms of productivity (quantity), content novelty, and optimism (positive emotions) in content. In particular, we intend to answer the following research questions:

- *Do the state-wide lockdowns impact the productivity of content creators?*
- *Do the state-wide lockdowns impact the novelty of content created by content creators?*
- *Do the state-wide lockdowns impact the optimism in content created by content creators?*

---

[1] https://www.cnbc.com/2020/08/24/tiktok-reveals-us-global-user-growth-numbers-for-first-time.html

To answer these research questions, we leverage a natural experiment resulted from the heterogeneous lockdown-related decisions taken by state governors in U.S. In particular, we identified a number of states that enforced identical state-wide lockdowns precisely on the same day; and we further identified a number of similar states that did not enforce any lockdowns during the same time frame. Given that the decision about enforcing a state-wide lockdown is not controlled by TikTok content creators, the differential lockdown policies create a natural experimental setting that would allow us to use a Difference-in-Differences (DiD) approach to examine the potential impacts of state-wide lockdowns on content creators' productivity as well as the novelty and optimism in their content. In addition to leveraging the lockdowns as exogenous shocks for econometric identification, we used machine learning-based natural language processing to construct measures of two key work outcomes (content novelty and optimism), and we conducted various additional analyses to examine the robustness of our results.

Based on the DiD analyses on a matched sample of content creators who reside in the lockdown versus non-lockdown states, our results show that content creators in states that enforced state-wide lockdowns posted significantly more content after the lockdown when compared to content creators from the non-lockdown states. This finding suggests that there is a productivity boost due to the physical restrictions caused by the lockdowns. However, our results also show that content novelty and content optimism from content creators in lockdown states (compared to similar content creators in non-lockdown states) both decrease after the enforcement of the lockdowns.

Our findings provide important implications regarding productivity in the context of remote work, particularly for emerging job categories such as independent content creators. In line with prior work on the productivity effects of work-from-home (Bloom et al. 2015; Choudhury et al. 2020), our findings suggest that the productivity of workers increase after they are restricted to work from home, perhaps because they find more time to allocate to their work when they work from their homes. However, other unintended consequences include the reduction of novelty and optimism in their work. Therefore, the implications of our study would inform decision makers to be aware of some of the unintended consequences of setting policies around work-from-home. In particular, our findings have important implications for embedding digital innovations in remote learning and remote work ecosystems. As digital innovators design new platforms to build resilience, they need to be aware of the consequences of remote work and design their platforms accordingly. That is, the design of the new platforms to aid remote work and remote education should not be based on the assumption that subjects would behave the same way they used to (before the disruption), but rather, they may behave differently due to the change in their physical environment. Therefore, our research is a call to revisit the assumptions about individuals' behaviors when designing new platforms based on digital innovation.

The organization of this manuscript is as follows. The next section reviews the current literature related to digital resilience and user-generated content (UGC) creation and presents the arguments leading to our hypotheses. We present our data and variables in the next section. We then describe the empirical model and present descriptive statistics along with the results of our analyses. We conclude by discussing our findings, the theoretical and practical implications of this study, as well as the limitations and potential extensions of this study.

## 2. Literature Review

### 2.1 Digital Resilience during Crises

There is a stream of IS literature on digital resilience, wherein most of the works focus on the context of life-threatening events such as natural disasters and terrorist attacks. During those unexpected large-scale crises, information technologies, with their high level of agility and responsiveness, have been shown as effective means to disseminate information and support decision-making (Park et al. 2015). For instance, oftentimes, an incident is initially reported by a witness with a mobile communication device, the report is rapidly distributed through social media services, and mainstream media involvement follows (Oh et al. 2013).

In particular, social media can play a critical role in propagating emergency information. For example, during the 2017 Storm Cindy in the U.S., a limited number of Twitter handles including those belonging to news and weather agencies, became the main source of information regarding the storm (Kim et al. 2018). Similarly, a case study reported by Yates and Paquette (2011) shows that during the 2010 Haitian earthquake, social media enabled effective knowledge management in a dynamic emergency environment through establishing coordination among various disaster relief agencies, resulting in transformation of knowledge in a way that it would be better utilized by individuals and decision makers (Yates and Paquette 2011).

IT artifacts not only can help build resilience during natural disasters, but they can also be used to cope with man-made emergency incidents such as shootings, building/structure fires, and criminal incidents (Han et al. 2015). Specifically, IT can be incorporated into an ecosystem and facilitate three sequential yet distinct operations: mobilization, situation assessment, and intervention (Yang et al. 2013). In a study of emergency management networks in metropolitan areas in Florida, Kapucu and Garayev (2012) find that IT utilization is positively related to the sustainability of emergency management networks and call for more investment in IT infrastructure. Researchers have also signified the role of IT artifacts in the design of emergency systems. For example, Chen et al. (2007) offer general design principles for building better emergency response management systems, and Chen et al. (2013) offer specific recommendations for systems that are used in fire-related extreme events.

As one of the most destructive crises in history, COVID-19 pandemic created an unprecedented challenge for individuals around the globe. It struck the world economy with the biggest shock since the second world war, followed by a slump in consumer spending, numerous business bankruptcies, and a temporary loss of 500 million jobs.[2] Among the studies that have examined the impact of the pandemic on the workforce, a significant number of them particularly focused on the disruptions that the lockdowns brought to academics. For instance, using a world-wide survey of researchers, Myers et al. (2020) find that the pandemic resulted in a sharp decline in time spent on academic research. A few other studies also reveal the presence of gender inequality in the change of research productivity during the pandemic (Andersen et al.

---

[2] https://www.economist.com/leaders/2020/10/08/the-pandemic-has-caused-the-worlds-economies-to-diverge.

2020; Cui et al. 2020; Kim and Patterson 2020). In conclusion, recent studies have signaled some of the potential impacts of the pandemic and the subsequent lockdowns on the way people perform their work.

In order to control the spread of the virus, many local and national governments, including several U.S. states, counties, and municipalities, issued lockdown orders and mandated their residents to stay at home for a specific period of time. In return, these lockdowns resulted in major impacts on the way individuals live, the way they work, the places where they spend their time, and the people they interact with (Ghose et al. 2020). An analysis of location data in New York City (Bakker et al. 2020) revealed that a direct result of the lockdown is that distance traveled every day dropped by 70%, the number of social contacts in certain places decreased by 93%, and the number of people staying home the whole day increased from 20% to 60%. Furthermore, Bick et al. (2020), by conducting a survey on 5,000 working age adults, show that 35.2 percent of the workforce worked entirely from home in May 2020, up from 8.2 percent in February 2020. In another study, Bartik et al. (2020) also find an increase in remote work. However, they find that the increase in remote work varies across industries. For example, remote work is more common in industries with more educated and higher-paid employees. And the productivity effect across industries vary with education.

As the pandemic and its impacts on society is evolving, a few pioneering studies have examined the individuals' use of digital platforms during the pandemic. For instance, Nabity-Grover et al. (2020) examine the pandemic-related self-disclosure behavior on social media. More specifically, they argue that the pandemic may have changed some topics from private to public knowledge (i.e., inside-out) and from readily shared to hidden (e.g., outside-in). Further, a recent study by Rao et al. (2020) analyzes tweets posted by government and health officials during the pandemic. The analysis of the tweets revealed that the officials often use either an alarming tone or a reassuring tone when they discuss COVID-19 topics on Twitter. The study suggests that, as the officials use Twitter to post alarming content, they also need to use a reassuring tone during the crisis (Rao et al. 2020). Overall, this emerging stream of literature has established the importance of social media in responding to crisis situations.

## 2.2 User-Generated Content (UGC) Creation

This study is closely related to UGC creation literature, which includes a variety of themes ranging from how to stimulate content creation (Burtch et al. 2018; Goes et al. 2016; Huang et al. 2019; Khern-am-nuai et al. 2018; Sun et al. 2017), to how to evoke novel content creation (Burtch et al. 2020), to examining various impacts of the emotions embedded in the content (Yin et al. 2014). These studies have largely expanded our understanding of mechanisms that would result in desired outcomes such as creating more content by users or creating specific types of content by them. In this section, we review the UGC literature related to content volume, content novelty, and content optimism, and further theorize the impacts of the change in individuals' physical environment (due to lockdown) on their content volume, content novelty, and content optimism.

### 2.2.1 Content Volume

The proliferation of Web 2.0 technology has resulted in a wide-range of UGC-centered platforms in a variety of use cases including purchase and investment decisions (Huang et al. 2019), consumer analytics (Lee and Bradlow 2011), and entertainment (Moon et al. 2014) to name a few. An essential element of UGC is related to users' motivations for content creation. IS research has identified several mechanisms to explain content creation. Sun et al. (2017), for example, examine the moderating effect of social connectedness (intrinsic motivation) on the relationship between monetary rewards and product review contribution. Burtch et al. (2018) reveal that the combination of financial incentives and social norms result in motivating users to write reviews in greater numbers and of greater length (Burtch et al. 2018). Design-based incentives are also shown to drive content creation. For instance, Goes et al. (2016) find evidence that incentive hierarchies, defined as achieving higher status within the platform, may temporarily motivate users to contribute more in a crowd-based online knowledge exchange. Additionally, gamification-based designs (symbolic awards) are found to have a sizeable effect on user contribution and retention (Gallus 2017). And recently, content creation for digital platforms has become a viable profession for many individuals (Tang et al. 2012). For example, on platforms such as YouTube or TikTok, content creators are able to earn revenue through advertising and other sources.

Despite the breadth of research in this area, there is a glaring lack of research in identifying the potential impacts of changes in users' physical environment on their content-creation behaviors. Filling this gap in the UGC literature is even more critical due to the fact that, as a result of the pandemic and the subsequent lockdowns, many individuals including UGC content creators have been mandated to work from home (WFH). Understanding the potential impacts of such a drastic change in individuals' physical environment on their productivity can inform the design of digital platforms in the hope of building a more resilient ecosystem. The main question is whether working from home resulted in a higher or a lower level of productivity. We argue that WFH facilitates greater work productivity, and we rationalize this argument in two ways:

First, individuals who work from home, on average, would have more time to spend on their work. For starters, those who work from home spend no time commuting to their workplaces. A study by Bloom (2015) also suggest that individuals who work from home have more time to spend on working because they do not run errands at lunch, they can start to work earlier, take shorter and fewer breaks, and end their work late in the day. Consequently, the increase in working hours can drive productivity higher. Second, people at home can have a relatively quiet environment and spend less time socializing with their colleagues. Workplaces could be perceived as distracting environments (Lee and Brand 2010) that could hinder individuals' focus and concentration (Roper and Juneja 2008). An empirical study using data from Ctrip, the largest travel agency in China, revealed that switching employees from in-office to WFH led to a 13.5% increase in employee productivity (Bloom et al. 2015). By conducting interviews with employees who worked from home, Bloom and colleagues attributed the increase of productivity to the quieter environment (home) and longer working time. Similarly, in our case, due to the physical restrictions enforced by the lockdowns, content creators who stayed at home, had more time to spend on content planning and shooting, video editing and production, and were less distracted by the surrounding environment or socializing with others. Therefore, we propose that they are likely to create more content:

*H1: Ceteris paribus, individuals mandated to stay at home create more content compared to those who are not mandated to stay at home.*

**2.2.2 Content Novelty**

Content novelty is an important element in UGC and has been shown to be a good predictor of content popularity (Carmel et al. 2012). Although originally rooted in psychology literature, content novelty has been an important construct in organization science (Hargadon and Bechky 2006; Sosa 2011) and marketing literature (Smith et al. 2007; Yang and Smith 2009). To gain a deeper understanding of the potential impact of the change in individuals' physical environment and the novelty of their content, we refer to the psychology literature. As suggested by Perry-Smith and Shalley (2003, *p*. 91), novelty is not a static feature, but rather is determined within "the bounds of social, cultural, and historical precedents of the field." In other words, while individuals are the sources of novelty, novelty stems from the interplay between the individuals and the environment around them (Cattani and Ferriani 2008). In the film industry, for instance, a longitudinal study finds that the socio-relational system nurtures individuals' creative performance (Cattani and Ferriani 2008). Woodman et al. (1993) proposed an interactionist model that nests individual creativity within groups, which "constitutes the social context in which the creative behavior occurs" (*p*. 303). Social interactions such as help-seeking and help-giving can catalyze creativity (Hargadon and Bechky 2006). Similarly, Sosa (2011) shows that having strong ties within a social network serves as an effective predictor for generating creative ideas. In summary, the evidence from prior studies commonly suggests that people need social interactions to generate new, novel, and creative ideas. While lockdowns can result in additional free time and a less distracting environment, at the same time, people interact with fewer friends, colleagues, and extended family members in person. The lack of in-person interactions, in turn, would give rise to an enhanced perception of social isolation (Bullinger et al. 2020). In fact, a survey study revealed that during the recent stay-at-home order in northern California, 56.4% of the participants reported perceived isolation and 36% of the participants reported loneliness (Gaeta and Brydges 2020). Taken together, we argue that the social isolation caused by the lockdowns can impact the level of content novelty in a negative way. In our case, individuals who are deprived of in-person interactions due to the state-wide lockdown orders are likely to create less novel content during the lockdowns. Therefore, we propose our second hypothesis:

*H2: Ceteris paribus, individuals mandated to stay at home create less novel content compared to those who are not mandated to stay at home.*

**2.2.3 Content Optimism**

While continuing work output and maintaining the quality of work (e.g., novelty) during the crises are both important, exhibiting optimism under the difficult time is another dimension of resilience (Connor and Davidson 2003). Since optimism would be reflected in positive emotions embedded in the content, content optimism in the context we study refers to the presence of positive emotions in UGC content.

Emotion embedded in content is an intriguing research topic that has drawn a great deal of attention. Many studies adopt text analysis or topic classification methods and apply machine learning-based sentiment analysis to measure the emotions embedded in textual documents. Related studies in information systems (IS) and marketing have advanced our understanding of emotions embedded in UGC. Broadly speaking, we can divide these studies into two streams; one stream of research examines emotion as an outcome, and the other stream examines emotion as an antecedent. To name a few studies from the first stream, Anderson and Agarwal (2011) demonstrated the significant role of emotion in one's decision-making process of health information disclosure. Huang et al. (2017) find that social network integration increases positive emotion and decreases negative emotion in the online review text. To name a few studies from the second stream, in the context of online reviews, studies find differential impact of anxiety and anger (Yin et al. 2014) in terms of perceived reviewer effort, as well as the relationship between the expressed arousal and review helpfulness (Yin et al. 2017). Research focusing on security reveals that positive emotion such as achievement and interest can encourage precaution-taking behaviors, while negative emotion inhibits precaution-taking behaviors (Burns et al. 2019). In terms of the application of emotion in other contexts such as healthcare, Wang et al. (2018) found that there is a curvilinear relationship between emotional social support and HIV patients' self-care behavior.

Similar to the gap we identified in the literature on content volume and content novelty, the bulk of the IS literature on content emotion has focused on the interplay between emotional content and IS constructs; with emotional content being either the outcome or the antecedent. Notably, there is a lack of research in connecting the change in the physical environment to one's emotional expressions. As noted in the previous section, one of the consequences of the lockdowns is the increased perception of social isolation, anxiety, and loneliness (Gaeta and Brydges 2020). Using Google trend data, Brodeur et al. (2020) indeed find a substantial increase in the search intensity for boredom in Europe and the U.S. during the lockdowns. They also find a significant increase in searches for loneliness, worry, and sadness. According to the self-determination theory established by Deci and Ryan (2000), relatedness, which refers to the desire to feel connected to others – to love and care and to be loved and cared for, is one of the innate psychological needs for human beings (Baumeister and Leary 1995; Deci and Ryan 2000). Thus, feeling related or connected to others is a part of eudemonic living that can foster well-being (Ryan et al. 2008). On the other hand, the feeling of loneliness can be negatively associated with optimism (Davis et al. 1992; Ben-Zur 2012). Given the above discussions, we expect to observe a drop in optimism in content created by individuals who were required to stay at their homes. Therefore, we propose our third hypothesis:

*H3: Ceteris paribus, individuals mandated to stay at home create less positive content compared to those who are not mandated to stay at home.*

# 3. Research Setting, Data, & Measures

## 3.1 Research Setting & Data

We collect data from TikTok, a short-video sharing social media platform that allows users to create short original video content and share those videos with other users on the mobile application called TikTok (Zhu et al. 2020). Founded in 2012 and launched in 2017 globally, it is the fastest growing social media platform in the world (Omar and Dequan 2020), and is now available in over 150 countries. As of October 2020, it is reported that TikTok has more than 800 million active users with almost 2 billion installations worldwide.[3] TikTok is different from other social media platforms due to its emphasis on sharing short videos, simple-to-use video editing, and music-inclusion functionalities (Chen et al. 2019). We are specifically interested in studying TikTok due to the fact that content creation in TikTok would require a more substantial effort when compared to other social media platforms such as Instagram, Twitter, and Facebook, and that the novelty of the content is an integral part of TikTok platform. In addition, career professionals in a variety of fields have been tapping TikTok as a platform to create and share their content.[4]

To examine the potential impact of lockdown orders, we first identified the treatment group and the control group. There were five states (Arkansas, Iowa, Nebraska, North Dakota, and South Dakota) that did not issue state-wide lockdown orders. All of the other states in the U.S. issued state-wide lockdown orders at some point.[5] The lockdown orders were issued over time from as early as March 19, 2020 (California) to as late as April 13, 2020 (Rhode Island).

---

[3] https://wallaroomedia.com/blog/social-media/tiktok-statistics/
[4] https://www.cnn.com/2020/11/27/health/tiktok-therapist-mental-health-intl-wellness/index.html and https://www.forbes.com/sites/forbesagencycouncil/2020/03/12/how-professionals-can-use-tiktok-as-a-marketing-channel/?sh=32cbd8114a02
[5] https://www.nbcnews.com/health/health-news/here-are-stay-home-orders-across-country-n1168736

Therefore, to form a relatively balanced treatment group, we targeted a date when there were, ideally, five states issuing lockdown orders. This would allow us to create comparable control and treatment groups, each including five states. We found that on March 23, 2020, five states announced their lockdown orders: Connecticut, Louisiana, Ohio, Oregon, and Washington. Thus, TikTok users residing in AR, IA, NE, ND, SD form our treatment group, and those who reside in CT, LA, OH, OR, and WA form our control group. Since there is no geographic information embedded in the short videos, we used video hashtags to identify user locations. Many users add a hashtag in the text description of the video to show their location (e.g., state). For example, in Figure 1, the user put the hashtag "#iowa" to reveal her/his location while watching the sunrise.

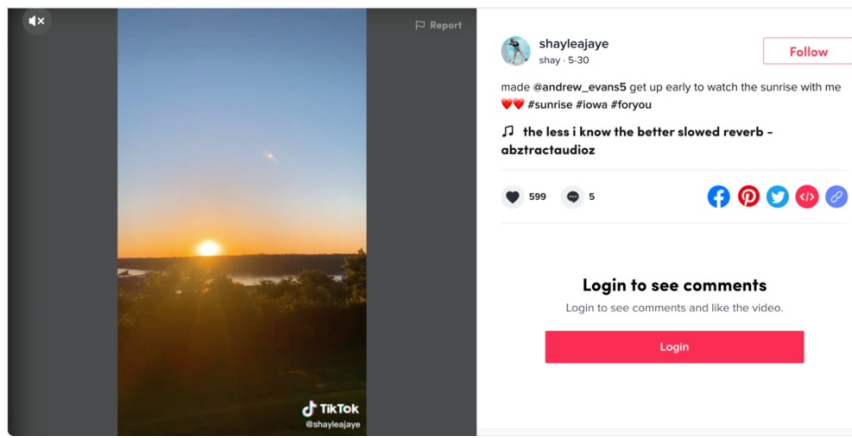

**Figure 1. An Example TikTok video with Hashtag "#iowa"**

Therefore, we obtained user information using ten hashtags representing ten states of interest, respectively: #arkansas, #iowa, #nebraska, #northdakota, #southdakota, #connecticut, #louisiana, #ohio, #oregon, and #washington. Due to the concern about the ambiguity in the location of users who use the hashtag "#washington" around whether it refers to DC or WA, we excluded the data from the state of Washington from our sample. It is worth noting that there is a fraction of users who post a hashtag containing a state but actually not residing in the state. For example, s/he may travel to another state and tag the state for the view. To this end, we carefully examined each user's location by reviewing her/his other videos and removed a small set of users who show inconsistencies in their user locations.

Once the users were identified, we collected information on the users and the short videos they published within the time period starting from November 1st, 2019 and ending to June 30th, 2020. Only the users who published at least one video during this time frame were included in our data. With the lockdown orders issued on March 23, 2020, we identified 21 weeks as the pre-treatment period and 14 weeks as the post-treatment period. The resulted data set contains 217,419 video observations that were posted by 2,346 unique TikTok users. On average, each user posted 92.7 videos during the time frame of our study.

**3.2 Variables & Measurements**

Our dependent variables of interest are content volume, content novelty, and content optimism. Content volume is measured by counting the weekly number of videos posted on the platform by a focal user. To construct measures of content novelty and content optimism, we leveraged methods and models in natural language processing, as detailed below:

When TikTok users post their short videos, they also provide a brief description of the video. We decided to use these descriptions to understand users' behaviors pre- and post-lockdown orders. To this end, we began with pre-processing steps to prepare the text descriptions for our analytical models. In the pre-processing steps, we first removed any html mark-ups such as http and https. Then we lemmatized the tokens and removed all of the stop-words using the spaCy package for Python.[6] These pre-processing steps resulted in clean text descriptions that we used as inputs in our text analytics. Below, we describe our approach in creating text-based measures for the two constructs we used in our study: content novelty and content optimism.

---

[6] We used spaCy's "en_core_web_lg" pretrained model. Documentations are available here: https://spacy.io/models/en

### 3.2.1. Measuring Content Novelty

Recent studies have used text-mining methods to identify novel ideas in text documents (Balsmeier et al. 2017; Toubia and Netzer 2017). One common approach in measuring novelty in UGC is to first create a vector representation of content and subsequently use those vector representations to calculate distance metrics such as Jaccard or cosine similarity (Hass 2017). For instance, Burtch et al. (2020) used this method to measure content novelty in Reddit and found that the recipients of peer awards will create content that is significantly similar to their own past content (i.e., less novel).

In this study, we use a similar approach by transforming the video descriptions to vector representations and then measure content novelty by comparing the vector representations of video descriptions created by a user with her/his previous content. If we convert the video descriptions to vectors, then we could use a measure of spatial distance to compute the differences between these vectors. For instance, if we have two vectors $u$ and $v$, we could use the following formula to compute their cosine distance:

$$cosine\_distance(u, v) = 1 - \frac{u \cdot v}{||u||_2 \, ||v||_2} \quad (1)$$

where $u \cdot v$ is the dot product of $u$ and $v$.

We define that if a user creates a novel video, it will contain something different from her/his previous videos. And a novel video would be accompanied by a novel description, and the novel description should be different from the previous descriptions created by that user. Therefore, the cosine distance between the novel video description ($u$) and the user's previous descriptions ($v$) should be large. On the other hand, if the new video description is not novel, then it would be similar to the previous descriptions created by the user and therefore should have a smaller value for cosine distance. As noted earlier, this approach for measuring novelty through text analytics is based on one proposed by Burtch et al. (2020). Given that we can measure the novelty of video descriptions using cosine distance, the key task we need to perform is to convert video descriptions to vectors with numeric values.

Vector representation of text documents is at the core of the majority natural language processing tasks. There are many different methods such as *Count Vectorizer*,[7] *TF-IDF Vectorizer*,[8] *Doc2Vec*,[9] and more recently, *Sentence-BERT*[10] for converting text sentences or documents into vector representations. Different from Burtch et al. (2020), which uses TF-IDF and Doc2Vec, we use an attention model, *Sentence-BERT*, which is based on Google's Bidirectional Encoder Representations from Transformers (BERT). BERT is currently considered the state-of-the-art model for many NLP tasks (Tenney et al. 2019). And BERT has been shown to be a very effective method for converting text documents to vector representations (Reimers and Gurevych 2019). Below, we explain how BERT creates representations of text data. Then, we explain how we used a modified BERT model to create vector representations of video descriptions.

BERT is an open-source language representation model introduced by Google in 2018. BERT is a pre-trained deep learning model that could be fine-tuned for a variety of NLP tasks and based on a variety of text data (Devlin et al. 2018). It uses Transformer, which is an attention mechanism that could learn contextual relationships between tokens (e.g., words) in a text document. Essentially, transformers are models that process words in relation to all the other words in a sentence, rather than one-by-one in order. BERT models can therefore consider the full context of a word by looking at the words that come before and after it.[11] Google has released two versions of BERT: BERT Base and BERT Large. Both models are developed based on deep neural networks. There are 12 hidden layers in BERT Base model while there are 24 hidden layers in BERT Large model. BERT is a pre-trained model, which means it has already been trained and built using text data. Google used two text data sets to train BERT: 2.5B words from Wikipedia and 800M words from BookCorpus (Devlin et al. 2018). BERT Base returns a vector representation of length 768 while BERT Large returns a vector representation of length 1028 for each token. These vector representations can then be used in NLP tasks such as text classification, question-answering, and natural language inference. The details of BERT, how it is trained, and how it can be used in NLP tasks are discussed in (Devlin et al. 2018) and its GitHub repository.[12]

---

[7] https://scikit-learn.org/stable/modules/generated/sklearn.feature_extraction.text.CountVectorizer.html
[8] https://scikit-learn.org/stable/modules/generated/sklearn.feature_extraction.text.TfidfVectorizer.html
[9] https://radimrehurek.com/gensim/models/doc2vec.html
[10] https://github.com/UKPLab/sentence-transformers

In our study, we were interested in obtaining a vector representation for each video description in our TikTok dataset. For this task, we chose BERT Base model, which returns a vector of length 768 for each token (e.g., word) in each sequence (e.g., sentence).[6] Each one of these vectors represents a token in the video description. To obtain the vector representation of the video descriptions (sequence of words), one could either take the mean of the vector representations of the tokens in the video description or obtain the vector representation of a special token that is automatically added to the beginning of each sequence before applying BERT. This token is called [CLS], and its vector representation can be used as the vector representation of the whole sequence (sentence). Although these two approaches are viable, research shows that they both result in poor sentence embeddings (Reimers and Gurevych 2019). Therefore, we decided to use a modified version of BERT that is designed for sentence embeddings. This model is called "Sentence-BERT" and is based on Siamese BERT networks (Reimers and Gurevych 2019).

---

[11] https://www.blog.google/products/search/search-language-understanding-bert.
[12] https://github.com/google-research/bert.

By applying Sentence-BERT to video descriptions, we obtained a vector of length 768 for each video description in our data. After obtaining the vector representations for each video description in our data set, given that our panel data is in weekly format, we aggregated these vector representations by taking their component-wise mean for each user for each week. To examine the quality of the aggregated vector representations, we deployed unsupervised machine learning algorithms t-distributed Stochastic Neighbor Embedding (t-SNE) and Balanced Iterative Reducing and Clustering using Hierarchies (BIRCH). We reported the details of this quality check and its results in Online Appendix A.

The aggregation resulted in a vector of length 768 for each user for each week the user posted any videos. Then, we measured cosine distance using specification (1), in which $u$ is the vector of component-wise averages of video descriptions posted by user $i$ during week $t$ and $v$ is the vector of component-wise averages of video descriptions posted by user $i$ during the last week before week $t$ that s/he posted any videos.[13]

### 3.2.2. Measuring Content Optimism

To measure content optimism (i.e., positive emotion) in video descriptions, we used Python package NRCLex.[14] This package is designed to measure emotional affect in text. NRCLex is a lexicon-based dictionary that uses a comprehensive list of words that are associated with specific emotional affects. NRCLex contains approximately 27,000 words, which are derived from the

---

[13] We used SciPy package's "spatial" module to measure the cosine distance. Documentations are available here: https://docs.scipy.org/doc/scipy/reference/generated/scipy.spatial.distance.cosine.html
[14] Documentations available in: https://pypi.org/project/NRCLex/

National Research Council Canada (NRC) affect lexicon[15] and the NLTK library's WordNet synonym sets.[16] We used function "affect_frequencies" from Python package NRCLex to obtain a score for positive affect in each video description in our data set.

It is worth noting that NRCLex provides a variety of other emotional affects, including fear, anger, anticipation, trust, surprise, sadness, disgust, and joy. For each video description, we obtained the scores for these additional emotional affects. These additional scores were used to create a matched sample of treated and control groups, as described in section 3.3. Furthermore, we used Python package "TextBlob"[17] to measure the level of subjectivity in video descriptions. The subjectivity score was also another construct that we used in section 3.3 to construct a matched sample of treated and control groups. The subjectivity score ranges from 0 (least subjective/ most objective) to 1 (most subjective/ least objective).

### 3.3 Matched Sample

The content creators included in the treatment group could systematically differ from those in the control group. Therefore, to make the samples in the two groups more comparable, we constructed matched samples before conducting the empirical analyses. We matched the samples at the user level; that is, for each user in the treatment group, we identified a similar user in the control group. In particular, we adopted the propensity score matching approach to balance the observed characteristics between the treatment and control groups. We calculated the propensity score using logit regression with an indicator of being treated (user in lockdown state) as a dichotomous outcome and a set of observed characteristics as covariates. The covariates primarily include the users' content characteristics prior to the treatment (lockdown), such as log-transformation of video count before the lockdown, video length (in seconds) before the lockdown, and a number of features extracted from the video descriptions such as Subjective, Fear, Anger, Anticipation, Trust, Surprise, Sadness, Disgust, Joy before the lockdown. Then, based on the propensity scores, we matched users between the treatment and control groups by applying the one-to-one nearest neighbor matching without replacement.

---

[15] Details are available in: https://nrc.canada.ca/en/research-development/products-services/technical-advisory-services/sentiment-emotion-lexicons
[16] Documentations available in: https://www.nltk.org/howto/wordnet.html
[17] Documentations available in: https://textblob.readthedocs.io/en/dev/

Table 1 presents the summary statistics of the treated and control groups before and after matching. The t-tests and p-values confirm that the means of the two groups are more similar after matching. Figure 2 presents the distribution of propensity scores for the treatment and control groups for both unmatched and matched samples. This figure indicates that the matched control group users have a propensity score distribution more similar to those in the treated group than those in the unmatched control group. These checks validate that the matching method is appropriate for producing similar groups.

**Table 1. Summary Statistics of Control and Treat Group Before and After Matching**

| Variable | Before Matching (N = 2346) | | | | After Matching (N= 2112) | | | |
|---|---|---|---|---|---|---|---|---|
| | Mean (Treated) | Mean (Control) | t | p-value | Mean (Treated) | Mean (Control) | t | p-value |
| Video Count | 0.82442 | 0.72328 | 3.52 | 0.000 | 0.82442 | 0.78117 | 1.44 | 0.150 |
| Video Length | 16.201 | 16.271 | -0.26 | 0.793 | 16.201 | 16.105 | 0.36 | 0.722 |
| Subjective | 0.23392 | 0.22383 | 1.54 | 0.124 | 0.23392 | 0.23175 | 0.32 | 0.748 |
| Fear | 0.07823 | 0.07689 | 1.15 | 0.248 | 0.07823 | 0.0781 | 0.11 | 0.909 |
| Anger | 0.07422 | 0.07243 | 1.65 | 0.099 | 0.07422 | 0.07362 | 0.55 | 0.583 |
| Anticipation | 0.07604 | 0.07478 | 1.05 | 0.293 | 0.07604 | 0.07584 | 0.17 | 0.868 |
| Trust | 0.08313 | 0.08285 | 0.17 | 0.866 | 0.08313 | 0.08231 | 0.56 | 0.574 |
| Surprise | 0.06896 | 0.06781 | 1.14 | 0.255 | 0.06896 | 0.06847 | 0.49 | 0.628 |
| Sadness | 0.07445 | 0.07342 | 0.96 | 0.335 | 0.07445 | 0.07414 | 0.29 | 0.772 |
| Disgust | 0.07587 | 0.07459 | 1.11 | 0.269 | 0.07587 | 0.07533 | 0.47 | 0.641 |
| Joy | 0.07894 | 0.07716 | 1.46 | 0.143 | 0.07894 | 0.0784 | 0.43 | 0.664 |

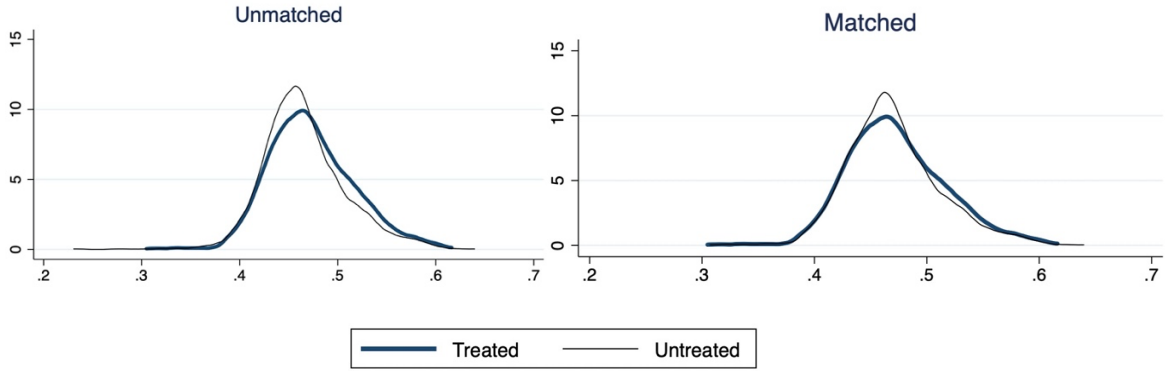

**Figure 2. Distribution of Propensity Scores for Treatment and Control Groups (Both Unmatched and Matched).**

## 4. Empirical Model and Results

### 4.1 Model-free Evidence

As a preliminary model-free evidence, Figure 3 visualizes the weekly trends in content volume from November 18, 2019 to June 30, 2020. The vertical line represents the week of March 23, 2020, which is the week the four states in the treatment group issued lockdown orders. We can observe that the gap in content volume between the two groups increases after the lockdown, suggesting an increase in content creation by individuals in lockdown states after the enforcement of lockdown.

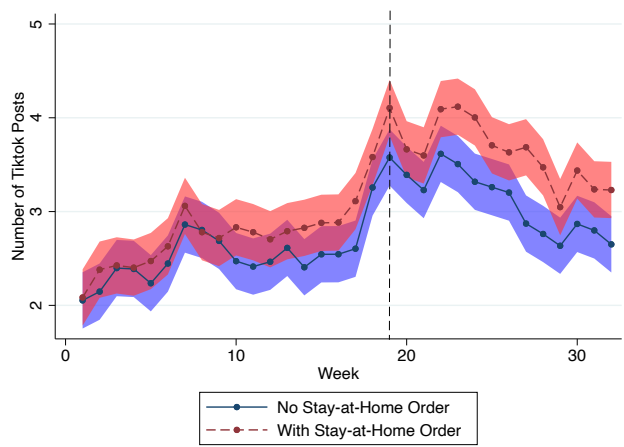

**Figure 3. Time Trends of the number of videos from November 2019 to June 2020**

## 4.2 Econometric Model Specification

After the outbreak of the coronavirus in the U.S., many states issued lockdown mandates between March 15 and April 15, 2020. As noted earlier, five states did not issue lockdown orders, resulting in a natural experiment setting that allows the comparison of the difference in the content creation behaviors before and after the lockdowns for states with the enforcement to the same difference for states that do not enforce lockdowns. Our identification exploits the lockdown as a result of the COVID-19 outbreak as an exogenous shock that has caused substantial disruptions in people's life by mandating residents to stay at home. The validity of our approach resides in the assumption that the shock is exogenous with respect to the users' anticipated responses. Given that the decisions about enforcing a state-wide lockdown are heterogeneous decisions taken by state governors, they are not controlled by TikTok content creators, ensuring the exogeneity of the shock. This identification strategy has been implemented in several prior studies such as Chan and Ghose (2014), Mousavi and Gu (2019), and Sun and Zhu (2013). To assess the effect of lockdown orders on content creation, we adopt the DiD methodology and specify the following model:

$$y_{it} = \beta_0 + \beta_1 \times Lockdown_t \times Treat_i + \gamma_t + \delta_i + \varepsilon_{it} \quad (2)$$

Where *i* denotes the user, *t* denotes the week, $\gamma_t$ is the time-fixed effect including a set of weekly time dummies that control for time trends, $\delta_i$ is the user-fixed effect that captures the time-invariant characteristics of user *i*, and $\varepsilon_{it}$ is the error term. The dummy variable *Lockdown* equals one if week *t* occurs after the lockdown was enforced (i.e., the week of March 23, 2020), and zero otherwise. Its main effect is absorbed by the time-fixed effects. *Treat* is a dummy for the treatment group (i.e., the states enforced lockdown order), which is set to one if the observations belong to the four states with lockdown order, and zero if the observations belong to the five states without lockdown order. Its main effect is absorbed by the user-fixed effects. We use specification (2) to study the impact of lockdown enforcement on three dependent variables ($y_{it}$). For the first dependent variable, content volume, we use the log-transformation of video count. In this case, the coefficient $β_1$ estimates the effect of lockdown on the content volume in lockdown states relative to the content volume in non-lockdown states after the lockdown enforcement. A positive and significant value for $β_1$ means the users generate more content (videos) in the lockdown states after the lockdown order. The second dependent variable is content novelty. In this case, the coefficient $β_1$ estimates the effect of lockdown on the content novelty in lockdown states relative to the content novelty in non-lockdown states after the lockdown order. A negative and significant value for $β_1$ means the content was less novel in the lockdown states after the lockdown order. The third dependent variable is content optimism, and the coefficient $β_1$ estimates the effect of lockdown on the content optimism embedded in the videos in lockdown states relative to that in non-lockdown states after the lockdown order. A negative and significant value for $β_1$ means the users express less optimism (less positive emotion) in the lockdown states after the lockdown order.

### 4.3 Main Results

Table 2 reports the estimated effect of the pandemic shock on content creation at the user level using Equation (2). First, the results in column 1 reveal a significant increase in content creators' productivity in terms of the volume of content that is produced. Specifically, we estimate that after the lockdown order, we see an increase of 7% in content volume in the lockdown states relative to the non-lockdown states. This result indicates that people who stay at home are more productive, supporting Hypothesis 1. With respect to content novelty, as reported in the second column of Table 2, the lockdown is negatively associated with the novelty of the content created by individuals in lockdown states relative to individuals in the non-lockdown states. Therefore, Hypothesis 2 is supported. Lastly, with respect to content optimism, our results reveal a post-lockdown reduction in the level of optimism expressed in video descriptions created by individuals in the lockdown states relative to individuals in non-lockdown states. Therefore, Hypothesis 3 is also supported. In addition to the main DiD models, we seek to conduct a series of falsification tests and robustness checks to rule out concerns of spurious correlations, which we report next.

Table 2. Impact of Lockdown on Content Creation

| Variable | (1) DV = Content Volume (log-transformed) | (2) DV = Content Novelty | (3) DV = Content Optimism |
|---|---|---|---|
| Post × Treat | 0.068** | -0.011* | -0.008* |
|  | (0.009) | (0.003) | (0.003) |
| Constant | 0.675*** | 0.309*** | 0.134*** |
|  | (0.013) | (0.004) | (0.004) |
| Week Fixed Effects | Yes | Yes | Yes |
| User Fixed Effects | Yes | Yes | Yes |
| # of Observations | 67,584 | 38,887 | 39,627 |
| # of User | 2,112 | 2,091 | 2,112 |
| Within R-squared | 0.025 | 0.005 | 0.002 |

**Notes:** Cluster-robust standard errors in parentheses; $* p < 0.05$; $** p < 0.01$; $*** p < 0.001$

## 4.4 Falsification Tests

### 4.4.1 Lockdowns' Impacts on Use of Common Words

Because there is no theoretical reason to expect that the pandemic shock would bear a relationship with the occurrence of common words, such as articles (e.g., the, a, an), filler expressions (e.g., you know, I mean), or numbers (e.g., first, hundred, thousand), we would not expect to observe a significant effect on these measures.[18] Thus, if we were to observe a significant effect, it would raise questions about the validity of our main results. As the results in Table 3 demonstrate, we observe no significant effects of lockdown on these outcome variables, lending further credence to our identification strategy.

---

[18] It is worth-noting that, for this analysis, we revised our previous pre-processing steps here to avoid removing stop-words.

Table 3. Impact of Lockdown on Common Words in Video Descriptions

| Variable | (1) DV = Article | (2) DV = Filler | (3) DV = Numbers |
|---|---|---|---|
| Post × Treat | 0.006 | 0.034 | 0.118 |
|  | (0.071) | (0.025) | (0.075) |
| Constant | 1.968*** | 0.055** | 1.080*** |
|  | (0.090) | (0.019) | (0.080) |
| Week Fixed Effects | Yes | Yes | Yes |
| User Fixed Effects | Yes | Yes | Yes |
| # of Observations | 39,627 | 39,627 | 39,627 |
| # of User | 2,112 | 2,112 | 2,112 |
| Within R-squared | 0.001 | 0.001 | 0.002 |

Notes: Cluster-robust standard errors in parentheses; $* \, p < 0.05$; $** \, p < 0.01$; $*** \, p < 0.001$

### 4.4.2 Randomization Inference

Another way to address serial correlation is to employ a randomization inference method (Bertrand et al. 2004). In this approach, computing the standard error for a specific experiment consists of a two-step approach. First, the DiD estimates for a large number of randomly generated placebo laws are estimated. Then the empirical distribution of the estimated effects for these placebo laws is used to form a significance test for the true law. In our case, we started by estimating the DiD estimate ($\beta_1$ in specification (2)) using the observed data. The next step is to randomly generate many placebo data and to run the model in specification (2) on these sets of placebo data. We created 10,000 placebo data sets. We then ran a model with specification (2) using each placebo set and obtained the corresponding DiD coefficient for each set. After running the model on all 10,000 placebo sets (and obtaining the 10,000 corresponding DiD estimates), we compared the actual DiD estimate obtained in the first step with the distribution of the placebo DiD estimates. We set the significance level at 0.05. Table 4 reports the results of this procedure for our three dependent variables. Based on the results, for all three dependent variables, the p-values are smaller than 0.05, which indicates that the DiD coefficient for the real data is statistically different from DiD coefficients obtained from placebo data. These results confirm our findings about the impacts of lockdowns on content volume, content novelty, and content optimism.

Table 4. Randomization Inference Results with 10,000 Simulations

| Variable | (1) DV = Content Volume (log-transformed) | (2) DV = Content Novelty | (3) DV = Content Optimism |
|---|---|---|---|
| P-value | **0.005** | **0.038** | **0.036** |
| User Fixed Effects | Yes | Yes | Yes |
| # of User | 2,112 | 2,112 | 2,112 |

**4.5 Robustness Checks**

To examine the robustness of our results, we conducted several additional robustness checks:

**4.5.1 Ignoring Time-Series Information**

Since we ran our analyses on weekly panel data, the estimated standard errors in our models may be serially correlated. This is especially problematic because the content creation itself is evidently serially correlated, which will exacerbate the bias in standard errors. To address this problem, we employ a remedy suggested by Bertrand et al. (2004). We collapse the time-series information into "pre-" and "post-" periods to eliminate the serial correlation observed in the weekly panel data. This strategy would tease out the potential effects of inconsistent standard errors due to serially correlated outcomes. To construct the collapsed dependent variables, we calculate simple average of the three outcome variables before lockdown ($y^{pre}$) and after lockdown ($y^{post}$). According to Table 5, the results in three models confirm our previous findings. The impact of lockdown on content volume is positive and significant, whereas its impact on content novelty and content optimism is negative and significant.

Table 5. Impact of Lockdown on Collapsed DVs

| Variable | (1) DV = Content Volume (log-transformed) | (2) DV = Content Novelty | (3) DV = Content Optimism |
|---|---|---|---|
| Post × Treat | **0.068**\*\* | -0.012\* | -0.011\* |
|  | (0.023) | (0.006) | (0.005) |
| Post | 0.097\*\*\* | -0.003 | 0.002 |
|  | (0.016) | (0.004) | (0.003) |
| Constant | 0.803\*\*\* | 0.350\*\*\* | 0.136\*\*\* |
|  | (0.008) | (0.001) | (0.001) |
| User Fixed Effects | Yes | Yes | Yes |
| # of Observations | 4,224 | 4,041 | 4,132 |
| # of User | 2,112 | 2,091 | 2,112 |
| R-squared | 0.06 | 0.01 | 0.004 |

**Notes:** Cluster-robust standard errors in parentheses; $*\, p < 0.05$; $**\, p < 0.01$; $***\, p < 0.001$

### 4.5.2 Parallel Trends Test

We test the parallel trends assumption by performing an analysis similar to the analysis in Cui et al. (2020), Seamans and Zhu (2013), Cui et al. (2019), and Calvo et al. (2019), where we expand specification (2) to estimate the treatment effect week by week before the shock. Specifically, we replace $Post_t$ in specification (2) with week dummy variables $WeekDummy_\tau^t$, where $\tau \in [-9, \ldots, -2, -1, 0]$ and $WeekDummy_\tau^t = 1$ if $\tau = t$ and 0 otherwise, indicating the relative $\tau$th week to the outbreak,

$$\ln(DV_{it}) = \beta_0 + \beta_1 \times \sum_{\tau=-9}^{-1} WeekDummy_\tau^t \times Treat_i + \sum_{\tau=-9}^{-1} WeekDummy_\tau^t + \delta_i + \varepsilon_{it} \quad (3)$$

The benchmark group is the week of the pandemic outbreak. The coefficients $\beta_{-9}$ to $\beta_{-1}$ identify any week-by-week pre-treatment difference between the states with lockdown orders and states without lockdown orders, which we expect to be insignificant. Table 6 reports the results of the parallel trends test. The test results show no pre-treatment differences in any DVs between control and treatment groups, which support the parallel trends assumption.

## Table 6. Pre-Treatment Trends

| Variable | (1) DV = Content Volume (log-transformed) | (2) DV = Content Novelty | (3) DV = Content Optimism |
|---|---|---|---|
| Treat × Week$_{-9}$ | -0.015 | 0.019 | 0.005 |
|  | (0.033) | (0.011) | (0.010) |
| Treat × Week$_{-8}$ | -0.023 | 0.005 | 0.005 |
|  | (0.033) | (0.011) | (0.010) |
| Treat × Week$_{-7}$ | -0.031 | 0.002 | 0.005 |
|  | (0.033) | (0.011) | (0.010) |
| Treat × Week$_{-6}$ | -0.042 | 0.004 | 0.021* |
|  | (0.033) | (0.010) | (0.010) |
| Treat × Week$_{-5}$ | -0.048 | 0.009 | 0.006 |
|  | (0.033) | (0.011) | (0.010) |
| Treat × Week$_{-4}$ | -0.044 | 0.001 | 0.013 |
|  | (0.033) | (0.010) | (0.010) |
| Treat × Week$_{-3}$ | -0.049 | 0.001 | 0.017 |
|  | (0.033) | (0.010) | (0.010) |
| Treat × Week$_{-2}$ | -0.039 | 0.005 | 0.008 |
|  | (0.033) | (0.010) | (0.010) |
| Treat × Week$_{-1}$ | -0.017 | 0.005 | 0.006 |
|  | (0.033) | (0.010) | (0.010) |
| Constant | 1.034*** | 0.271*** | 0.134*** |
|  | (0.012) | (0.004) | (0.003) |
| Week Fixed Effects | Yes | Yes | Yes |
| User Fixed Effects | Yes | Yes | Yes |
| # of Observations | 23,232 | 12,778 | 13,349 |
| # of Users | 2,112 | 1,907 | 2,022 |
| Within R-squared | 0.03 | 0.03 | 0.003 |

**Notes:** Cluster-robust standard errors in parentheses; $* p < 0.05; ** p < 0.01; *** p < 0.001$

### 4.5.3 Effective Lockdown Period

When the states with lockdown orders issued one, many of them also specified an end date for the order. In particular, three of the four states in our treatment group had an end date: lockdown order issued by Connecticut's ended on May 20, Louisiana's lockdown ended on May 15, and Ohio's lockdown ended on May 1. Considering that individuals may change their behaviors after the expiration of the lockdowns, which may dilute our results, we re-run our main model using only the effective lockdown periods stated above. Table 7 reports the results of this test, which are consistent with the results in our main model.

**Table 7. Impact of Lockdowns on Content Creation during Effective Lockdown Periods**

| Variable | (1) DV = Content Volume (log-transformed) | (2) DV = Content Novelty | (3) DV = Content Optimism |
|---|---|---|---|
| Post × Treat | **0.077**** | **-0.014*** | **-0.009*** |
|  | (0.024) | (0.005) | (0.004) |
| Constant | 0.672*** | 0.311*** | 0.134*** |
|  | (0.014) | (0.005) | (0.004) |
| Week Fixed Effects | Yes | Yes | Yes |
| User Fixed Effects | Yes | Yes | Yes |
| # of Observations | 62,136 | 35,606 | 36,346 |
| # of User | 2,112 | 2,085 | 2,112 |
| R-squared | 0.03 | 0.01 | 0.002 |

**Notes:** Cluster-robust standard errors in parentheses; $* p < 0.05$; $** p < 0.01$; $*** p < 0.001$

## 5. Discussions

The unprecedented COVID-19 pandemic has significantly transformed the way individuals live and work (Ghose et al. 2020). During the outbreak of this pandemic, many countries around the world enforced nation-wide lockdown orders to mandate people to stay at their homes as much as possible. Many states in the U.S. also issued lockdown orders as the virus started to spread in the states. In this study, we examine how the lockdown orders, which forced individuals work from home, impact the content creators' work outcomes in terms of content volume, content novelty, and content optimism. To study this topic, we leveraged a natural experiment wherein four states issued lockdown orders on the same day, and five states never issued any lockdown orders. Using a unique dataset collected from TikTok, a short-video sharing social media platform, we combined econometric methods (difference-in-differences estimations of a matched sample) with machine learning-based natural language processing to find that compared to the individuals residing in the non-lockdown states, the individuals who work from home (i.e., residing in the lockdown states) produce more content after the lockdown orders. However, the novelty and optimism of the content created by the individuals in lockdown states decreased when compared to content created by individuals in non-lockdown states.

## 5.1 Theoretical Implications

First and foremost, this work extends the literature on digital resilience, i.e., the use of IT/IS in disaster management (Park et al. 2015; Rao et al. 2020; Yates and Paquette 2011). Although prior work reveals that the IT artifacts such as social media can be effectively used for information dissemination (Kim et al. 2018) and real-time alert provisions (Yates and Paquette 2011), our work explores the impact of the crisis-caused physical changes in users' work environment on their content creation behaviors in terms of the content volume, content novelty, and content optimism. While the pandemic and the subsequent stay-at-home orders have affected offline work significantly, the work enabled by digital platforms (e.g., creative content creation) does not rely on commuting to traditional offices and can be continued by individuals with more flexibility during the crisis. Our results suggest that individuals, at least to some extent, can cope with the lockdown as they produce a higher volume of content. However, at the same time, the level of content novelty and content optimism decreases, signaling an adverse effect on content quality. These findings advance our understanding of the work outcome responses to the pandemic in the space of digital creative content generation.

Second, our study contributes to the nascent COVID-19 literature in social science, which has largely focused on a) the COVID's influence on work productivity (Myers et al. 2020), particularly gender inequality in research productivity (Andersen et al. 2020; Cui et al. 2020); b) general descriptive studies reporting civilians' reactions to the pandemic (Andersen 2020; Coven and Gupta 2020); and c) theory-based studies such as research examining the trade-offs between privacy and social good (Ghose et al. 2020) and the role of individualism during crises (Bian et al. 2020). Our work explores how individuals react to the change in their physical environment with respect to their online content creation behavior. We are aware of only one study that examines the use of social media during the COVID-19 pandemic (Rao et al. 2020). The researchers analyze tweets produced by official accounts and examine how alarming messages due to retweets of officials' messages may vary in comparison to reassurance message retweets. They suggest that providing reassurance during the crisis is paramount for government officials, health experts, and national news media (Rao et al. 2020). While this research is a pioneer that studies people's response to the pandemic using social media, it is mainly exploratory, it discusses the retweeting behavior in general, and it does not reach induvial-level analysis. By exploiting a natural experiment caused by the pandemic and the subsequent state-wide lockdowns, our paper is the first attempt to examine online users' responses to the COVID-19 pandemic, how do they adjust themselves, and how this adjustment impacts the pattern of their responses.

Third, our study contributes to UGC literature (Goes et al. 2016; Huang et al. 2017; Khern-am-nuai et al. 2018; Yin et al. 2014) by studying the impact of a change in users' physical environment on their content creation behavior. More specifically, our work extends research on three important outcomes in content creation literature: content volume, content novelty, and content optimism. While previous works on these areas have identified various mechanisms to stimulate a larger volume of contents (Burtch et al. 2018), examined how to invoke novel content (Burtch et al. 2020; Gallus 2017), and studied the embedded emotion elements in the context of online reviews (Yin et al. 2014), they mainly conduct investigations within systems and examine the interplays between the UGC characteristics and system features. We contribute to this literature by studying the association between changes in individuals' physical environment and their content creation behaviors.

## 5.2 Managerial Implications

This study also has important managerial implications, particularly for the emerging job categories such as independent content creators. In line with some prior work on the productivity effects of work-from-home (Bloom et al. 2015; Choudhury et al. 2020), our findings reveal a similar positive impact on content creation as individuals who are mandated to work from their homes would find more time to spend on their work. However, the unintended consequences of working from home include the reduction of novelty and optimism in the content. These findings inform decision makers to be aware of some of the unintended consequences of setting policies around work-from-home. We believe that this study is important not just because it is an attempt to understand work-from-home behavior during lockdowns, but mainly because some organizations have already decided to let their employees work from home indefinitely (Bartik et al. 2020). The most prominent one, perhaps, is Microsoft that announced its decision in the midst of the pandemic.[19] In addition to Microsoft, a wide array of other companies have also decided to let at least a portion of their employees work from home in the long run.[20] Work-from-home has become such an integral part of our society that the Federal Reserve's chair, Jerome Powell, urged policymakers to brace it and to consider it as a trend posed to stay even after we put the COVID-19 pandemic behind us.[21] Hence, understanding the potential impacts of work-from-home positions on workers' productivity is important. As suggested by our findings, although individuals are productive, they exhibit less novelty and optimism in their contents. Therefore, it might be the case that employees are keeping their productivity level, but the quality of their work is compromised. This could cause a serious concern for the technology sector in which many jobs can be conducted remotely yet would require a significant level of novelty. Thus, firms are suggested to evaluate both the quantity and quality of their employees' work before they make a decision about whether to allow remote work in the long run. Furthermore, if employees work from home all the time, the lack of in-person communications with colleagues may jeopardize their mental well-being, as suggested by our result regarding the reduced optimism in content created by those who worked from home.

---

[19] https://www.forbes.com/sites/carlypage/2020/10/09/microsoft-will-let-employees-work-from-home-permanently/?sh=768c65a7172a
[20] https://www.businessinsider.com/companies-asking-employees-to-work-from-home-due-to-coronavirus-2020
[21] https://www.cnbc.com/2020/11/12/here-are-the-things-that-scare-jerome-powell-the-most-about-the-economy-right-now.html

In addition, our findings have important implications for embedding digital innovations in remote learning and remote work ecosystems. As digital innovators design new platforms to build resilience, they need to be aware of the consequences of remote work and design their platforms accordingly. That is, the design of the new platforms to aid remote work and remote education should be based on the assumption that subjects would behave differently as a result of the change in their physical environment. Therefore, our research is a call to revisit the assumptions about subjects' behaviors when designing new platforms based on digital innovation.

**5.3 Limitations and Future Research**

As with most empirical studies that rely on observational data, our study is not free of limitations. One of the limitations of our study is related to the sample. We identified users' locations based on the hashtags they used in their video descriptions or based on whether they revealed their locations in their user accounts. Therefore, this sampling strategy removes users who did not use a location hashtag or did not disclose their location in their user account from our sample. Therefore, future studies could conduct a larger-scale, more comprehensive study if researchers are able to obtain all users' location data. For example, future research could potentially partner with social media platforms and retrieve users' location information from their servers given users' consent.

Another limitation of our study is related to potential generalizability given our research context. The individuals in our study are creative content workers in social media platforms as opposed to workers in conventional jobs. Employees in conventional jobs are paid to perform specific tasks. Therefore, 'novelty' may be less relevant for some more routine, less creative kinds of jobs. And our findings likely will generalize to gig workers who engage in creative tasks, such as website design, content creation, advertising campaigns, and so on. Further, intrinsic motivation and extrinsic motivation could also take effects when employees perform their work in terms of productivity and quality/level of novelty. To this end, managers should employ a longer period of observation and comprehensive evaluation to decide whether they should allow employees to work remotely in the long run.

Lastly, while we were able to investigate the direct work outcomes (e.g., content volume, novelty, and optimism) during the lockdowns, we were not able to observe other important outcomes such as individuals' mental health status, which perhaps is one of the most important questions to answer to better understand digital resilience during crises. Therefore, while the current research makes the first effort to study individuals' reaction to the change in their physical work environment, it opens up many opportunities for future studies to better understand the individuals' work-related outcomes and psychological state in a changed physical environment or during a crisis.

# Reference

Andersen, J. P., Nielsen, M. W., Simone, N. L., Lewiss, R. E., and Jagsi, R. 2020. "Meta-Research: Is Covid-19 Amplifying the Authorship Gender Gap in the Medical Literature?" *Working Paper*. arXiv preprint arXiv:2005.06303.

Andersen, M. 2020. "Early Evidence on Social Distancing in Response to Covid-19 in the United States." *Working Paper*. Available at SSRN: https://ssrn.com/abstract=3569368 or http://dx.doi.org/10.2139/ssrn.3569368.

Anderson, C. L., and Agarwal, R. 2011. "The Digitization of Healthcare: Boundary Risks, Emotion, and Consumer Willingness to Disclose Personal Health Information," *Information Systems Research* (22:3), pp. 469-490.

Bakker, M., Berke, A., Groh, M., Pentland, A. S., and Moro, E. 2020. "Effect of Social Distancing Measures in the New York City Metropolitan Area." *Working Paper*. MIT

Balsmeier, B., Fleming, L., and Manso, G. 2017. "Independent Boards and Innovation," *Journal of Financial Economics* (123:3), pp. 536-557.

Bartik, A. W., Cullen, Z., Glaeser, E. L., Luca, M., and Stanton, C. 2020. "What Jobs Are Being Done at Home During the Covid-19 Crisis? Evidence from Firm-Level Surveys." *Working Paper*. Harvard Business School.

Baumeister, R. F., and Leary, M. R. 1995. "The Need to Belong: Desire for Interpersonal Attachments as a Fundamental Human Motivation," *Psychological Bulletin* (117:3), pp. 497.

Ben-Zur, H. 2012. "Loneliness, optimism, and well-being among married, divorced, and widowed individuals." *The Journal of Psychology* (146:1-2), pp. 23-36.

Bian, B., Li, J., Xu, T., and Foutz, N. Z. 2020. "Individualism During Crises." *Working Paper*. Available at SSRN: https://ssrn.com/abstract=3626841 or http://dx.doi.org/10.2139/ssrn.3626841.

Bick, A., Blandin, A., and Mertens, K. 2020. "Work from Home after the Covid-19 Outbreak." *Working Paper*. Available at SSRN: https://ssrn.com/abstract=3638737 or http://dx.doi.org/10.24149/wp2017.

Bloom, N., Liang, J., Roberts, J., & Ying, Z. J. 2015. Does working from home work? Evidence from a Chinese experiment. *The Quarterly Journal of Economics* (130:1), 165-218.

Brodeur, A., Clark, A. E., Fleche, S., and Powdthavee, N. 2020. "Assessing the Impact of the Coronavirus Lockdown on Unhappiness, Loneliness, and Boredom Using Google Trends." *Working Paper*. arXiv preprint arXiv:2004.12129.

Bullinger, L. R., Carr, J. B., and Packham, A. 2020. "Covid-19 and Crime: Effects of Stay-at-Home Orders on Domestic Violence." *NBER Working Paper No. 27667*.

Burns, A., Roberts, T. L., Posey, C., and Lowry, P. B. 2019. "The Adaptive Roles of Positive and Negative Emotions in Organizational Insiders' Security-Based Precaution Taking," *Information Systems Research* (30:4), pp. 1228-1247.

Burtch, G., He, Q., Hong, Y., and Lee, D. 2020. "Peer Awards Increase User Content Generation but Reduce Content Novelty ". *Working Paper*. Available at SSRN: https://ssrn.com/abstract=3465879 or http://dx.doi.org/10.2139/ssrn.3465879.

Burtch, G., Hong, Y., Bapna, R., and Griskevicius, V. 2018. "Stimulating Online Reviews by Combining Financial Incentives and Social Norms," *Management Science* (64:5), pp. 2065-2082.

Carmel, D., Roitman, H. and Yom-Tov, E. 2012. On the relationship between novelty and popularity of user-generated content. *ACM Transactions on Intelligent Systems and Technology* (3:4), pp. 1-19.

Cattani, G., and Ferriani, S. 2008. "A Core/Periphery Perspective on Individual Creative Performance: Social Networks and Cinematic Achievements in the Hollywood Film Industry," *Organization Science* (19:6), pp. 824-844.

Chen, R., Sharman, R., Rao, H.R. and Upadhyaya, S.J. 2007. Design principles for emergency response management systems. *Journal of Information Systems and e-Business Management* (5:3), pp. 81-98.

Chen, R., Sharman, R., Rao, H.R. and Upadhyaya, S.J. 2013. Data model development for fire related extreme events: An activity theory approach. *MIS Quarterly* (37:1), pp. 125-147.

Chen, Z., He, Q., Mao, Z., Chung, H.-M., and Maharjan, S. 2019. "A Study on the Characteristics of Douyin Short Videos and Implications for Edge Caching." *Working Paper*. http://arxiv.org/abs/1903.12399

Choudhury, P., Foroughi, C., & Larson, B. Z. 2020. Work-from-anywhere: The productivity effects of geographic flexibility. *Academy of Management Proceedings* (2020:1), pp. 21199. Briarcliff Manor, NY: Academy of Management.

Connor, K. M., and Davidson, J. R. 2003. Development of a new resilience scale: The Connor‐Davidson resilience scale (CD‐RISC). *Depression and Anxiety* (19:6), pp. 76-82.

Coven, J., and Gupta, A. 2020. "Disparities in Mobility Responses to Covid-19." *Working Paper*. NYU Stern School of Business.

Cui, R., Ding, H., and Zhu, F. 2020. "Gender Inequality in Research Productivity During the Covid-19 Pandemic." *Working Paper*. Available at SSRN: https://ssrn.com/abstract=3623492 or http://dx.doi.org/10.2139/ssrn.3623492.

Davis, S.F., Miller, K.M., Johnson, D., McAuley, K. and Dinges, D. 1992. The relationship between optimism-pessimism, loneliness, and death anxiety. *Bulletin of the Psychonomic Society* (30:2), pp. 135-136.

Deci, E. L., and Ryan, R. M. 2000. "The" What" and" Why" of Goal Pursuits: Human Needs and the Self-Determination of Behavior," *Psychological Inquiry* (11:4), pp. 227-268.

Devlin, J., Chang, M.-W., Lee, K., and Toutanova, K. 2018. BERT: Pre-Training of Deep Bidirectional Transformers for Language Understanding. *Working Paper*. http://arxiv.org/abs/1810.04805.

Fosslien, L., and Duffy, M. W. (2020). How to Combat Zoom Fatigue. *Harvard Business Review* https://hbr.org/2020/04/how-to-combat-zoom-fatigue.

Gaeta, L., and Brydges, C. R. 2020. "Coronavirus-Related Anxiety, Social Isolation, and Loneliness in Older Adults in Northern California During the Stay-at-Home Order," *Journal of Aging & Social Policy* (Forthcoming).

Gallus, J. 2017. "Fostering Public Good Contributions with Symbolic Awards: A Large-Scale Natural Field Experiment at Wikipedia," *Management Science* (63:12), pp. 3999-4015.

Ghose, A., Li, B., Macha, M., Sun, C., and Foutz, N. Z. 2020. "Trading Privacy for the Greater Social Good: How Did America React During Covid-19? ." *Working Paper*. Available at SSRN: https://ssrn.com/abstract=3624069 or http://dx.doi.org/10.2139/ssrn.3624069.

Goes, P. B., Guo, C., and Lin, M. 2016. "Do Incentive Hierarchies Induce User Effort? Evidence from an Online Knowledge Exchange," *Information Systems Research* (27:3), pp. 497-516.

Han, W., Ada, S., Sharman, R., & Rao, H. R. 2015. Campus emergency notification systems: An examination of factors affecting compliance with alerts. *MIS Quarterly* (39:4), 909-930.

Hargadon, A. B., and Bechky, B. A. 2006. "When Collections of Creatives Become Creative Collectives: A Field Study of Problem Solving at Work," *Organization Science* (17:4), pp. 484-500.

Hass, R. W. 2017. "Tracking the Dynamics of Divergent Thinking Via Semantic Distance: Analytic Methods and Theoretical Implications," *Memory & Cognition* (45:2), pp. 233-244.

Huang, N., Burtch, G., Gu, B., Hong, Y., Liang, C., Wang, K., Fu, D., and Yang, B. 2019. "Motivating User-Generated Content with Performance Feedback: Evidence from Randomized Field Experiments," *Management Science* (65:1), pp. 327-345.

Huang, N., Burtch, G., Hong, Y., Pavlou, PA. 2020. Unemployment and Worker Participation in the Gig Economy: Evidence from An Online Labor Market, *Information Systems Research*, (31:2), pp. 431-448.

Huang, N., Hong, Y., and Burtch, G. 2017. "Social Network Integration and User Content Generation: Evidence from Natural Experiments," *MIS Quarterly* (41:4), pp. 1035-1058.

Kapucu, N. and Garayev, V. 2013. Designing, managing, and sustaining functionally collaborative emergency management networks. *The American Review of Public Administration* (43:3), pp. 312-330.

Khern-am-nuai, W., Kannan, K., and Ghasemkhani, H. 2018. "Extrinsic Versus Intrinsic Rewards for Contributing Reviews in an Online Platform," *Information Systems Research* (29:4), pp. 871-892.

Kim, J., Bae, J., and Hastak, M. 2018. "Emergency Information Diffusion on Online Social Media During Storm Cindy in Us," *International Journal of Information Management* (40), pp. 153-165.

Lee, S. Y., and Brand, J. 2010. "Can Personal Control over the Physical Environment Ease Distractions in Office Workplaces?," *Ergonomics* (53:3), pp. 324-335.

Lee, T. Y., and Bradlow, E. T. 2011. "Automated Marketing Research Using Online Customer Reviews," *Journal of Marketing Research* (48:5), pp. 881-894.

Moon, Y. J., Kim, W. G., and Armstrong, D. J. 2014. "Exploring Neuroticism and Extraversion in Flow and User Generated Content Consumption," *Information & Management* (51:3), pp. 347-358.

Myers, K. R., Wei Yang Tham, Yian Yin, Nina Cohodes, Jerry G Thursby, Marie C Thursby, Peter E Schiffer, Joseph T Walsh, Karim R Lakhani, and Wang, D. 2020. "Quantifying the Immediate Effects of the Covid-19 Pandemic on Scientists." *Working Paper*. arXiv preprint arXiv:2005.11358.

Nabity-Grover, T., Cheung, C.M. and Thatcher, J.B. 2020. Inside out and outside in: How the COVID-19 pandemic affects self-disclosure on social media. *International Journal of Information Management* (55), pp. 102188.

Oh, O., Agrawal, M., and Rao, H. R. 2013. "Community Intelligence and Social Media Services: A Rumor Theoretic Analysis of Tweets During Social Crises," *MIS Quarterly* (37:2), pp. 407-426.

Omar, B., and Dequan, W. 2020. "Watch, Share or Create: The Influence of Personality Traits and User Motivation on Tiktok Mobile Video Usage," *International Journal of Interactive Mobile Technologies* (14:4), pp. 121-137.

Park, I., Sharman, R., and Rao, H. R. 2015. "Disaster Experience and Hospital Information Systems: An Examination of Perceived Information Assurance, Risk, Resilience, and His Usefulness," *MIS Quarterly* (39:2), pp. 317-344.

Perry-Smith, J. E., and Shalley, C. E. 2003. "The Social Side of Creativity: A Static and Dynamic Social Network Perspective," *Academy of Management Review* (28:1), pp. 89-106.

Rao, H. R., Vemprala, N., Akello, P., and Valecha, R. 2020. "Retweets of Officials' Alarming Vs Reassuring Messages During the Covid-19 Pandemic: Implications for Crisis Management," *International Journal of Information Management* (55), p. 102187.

Reimers, N., and Gurevych, I. 2019. "Sentence-BERT: Sentence Embeddings Using Siamese BERT-Networks," *Working Paper*. http://arxiv.org/abs/1908.10084.

Roper, K. O., and Juneja, P. 2008. "Distractions in the Workplace Revisited," *Journal of Facilities Management* (6:2), pp. 91-109.

Ryan, R. M., Huta, V., and Deci, E. L. 2008. "Living Well: A Self-Determination Theory Perspective on Eudaimonia," *Journal of Happiness Studies* (9:1), pp. 139-170.

Sosa, M. E. 2011. "Where Do Creative Interactions Come From? The Role of Tie Content and Social Networks," *Organization Science* (22:1), pp. 1-21.

Sun, Y., Dong, X., and McIntyre, S. 2017. "Motivation of User-Generated Content: Social Connectedness Moderates the Effects of Monetary Rewards," *Marketing Science* (36:3), pp. 329-337.

Tang, Q., Gu, B. and Whinston, A.B. 2012. Content contribution for revenue sharing and reputation in social media: A dynamic structural model. *Journal of Management Information Systems*, (29:2), pp. 41-76.

Tenney, I., Das, D., and Pavlick, E. 2019. BERT Rediscovers the Classical NLP. *Working Paper*. http://arxiv.org/abs/1905.05950.

Toubia, O., and Netzer, O. 2017. "Idea Generation, Creativity, and Prototypicality," *Marketing Science* (36:1), pp. 1-20.

Wang, X., Parameswaran, S., Bagul, D. M., and Kishore, R. 2018. "Can Online Social Support Be Detrimental in Stigmatized Chronic Diseases? A Quadratic Model of the Effects of Informational and Emotional Support on Self-Care Behavior of Hiv Patients," *Journal of the American Medical Informatics Association* (25:8), pp. 931-944.

Woodman, R. W., Sawyer, J. E., and Griffin, R. W. 1993. "Toward a Theory of Organizational Creativity," *Academy of Management Review* (18:2), pp. 293-321.

Yang, L., Yang, S.H. and Plotnick, L., 2013. How the internet of things technology enhances emergency response operations. *Technological Forecasting and Social Change* (80:9), pp.1854-1867.

Yates, D., and Paquette, S. 2011. "Emergency Knowledge Management and Social Media Technologies: A Case Study of the 2010 Haitian Earthquake," *International Journal of Information Management* (31:1), pp. 6-13.

Yin, D., Bond, S. D., and Zhang, H. 2014. "Anxious or Angry? Effects of Discrete Emotions on the Perceived Helpfulness of Online Reviews," *MIS Quarterly* (38:2), pp. 539-560.

Yin, D., Bond, S. D., and Zhang, H. 2017. "Keep Your Cool or Let It Out: Nonlinear Effects of Expressed Arousal on Perceptions of Consumer Reviews," *Journal of Marketing Research* (54:3), pp. 447-463.

Zhu, C., Xu, X., Zhang, W., Chen, J., and Evans, R. 2020. "How Health Communication Via Tik Tok Makes a Difference: A Content Analysis of Tik Tok Accounts Run by Chinese Provincial Health Committees," *International Journal of Environmental Research and Public Health* (17:1), pp. 192-205.

# Supplementary Online Appendix

**Appendix A: Examining the Quality of Sentence-BERT Embeddings**

To evaluate the quality of the embeddings, we decided to use those embeddings to cluster video descriptions. If the video descriptions within each cluster are related to each other, yet different from the video descriptions in other clusters, this could indicate that the embeddings helped us identify and recover the clusters properly. Given that the length of each embedding (vector representation) is 768, we use t-distributed Stochastic Neighbor Embedding (t-SNE) to reduce the dimensionality of the vector representation before cluster analysis. In t-SNE, the algorithm begins with converting similarities between data points to joint probabilities. Then, the algorithm tries to minimize the Kullback-Leibler divergence between the joint probabilities of the high-dimensional data and the low-dimensional embedding (Van Der Maaten and Hinton 2008). We used scikit-learn's TSNE function[22] to create a low-dimensional representation of the embeddings. Then, we used Python package "clusteval"[23] to determine the best value for the number of clusters. "clusteval" iteratively examines different values of k (number of clusters) to find the value with the best outcome. There are three different evaluation metrics implemented in "clusteval": 1- Silhouette, 2- Derivative, and 3- DBSCAN. Figure A.1 reports the result of this evaluation using Silhouette as the evaluation metric. The red dashed line in Figure A.1 corresponds to 6 as the best value for the number of clusters. That is, per this analysis, setting the number of clusters to 6 would result in the highest value for Silhouette (0.324). Therefore, we ran the BIRCH clustering algorithm by setting the value for the number of clusters to 6. Figure A.2 visualizes the clusters of video descriptions in our data set. Clusters are identified using different colors. The x axis is the first dimension of t-SNE and the y axis is the second dimension of t-SNE. As can be observed in Figure A.2, the clusters that are identified using BIRCH are quite distinct. That is, the data points in each cluster are closer to other data points of the same cluster than other clusters. This analysis signals that Sentence-BERT embedding was successful in recovering the structure in the video descriptions.

---

[22] Documentations available in https://scikit-learn.org/stable/modules/generated/sklearn.manifold.TSNE.html
[23] Documentations available in https://github.com/erdogant/clusteval

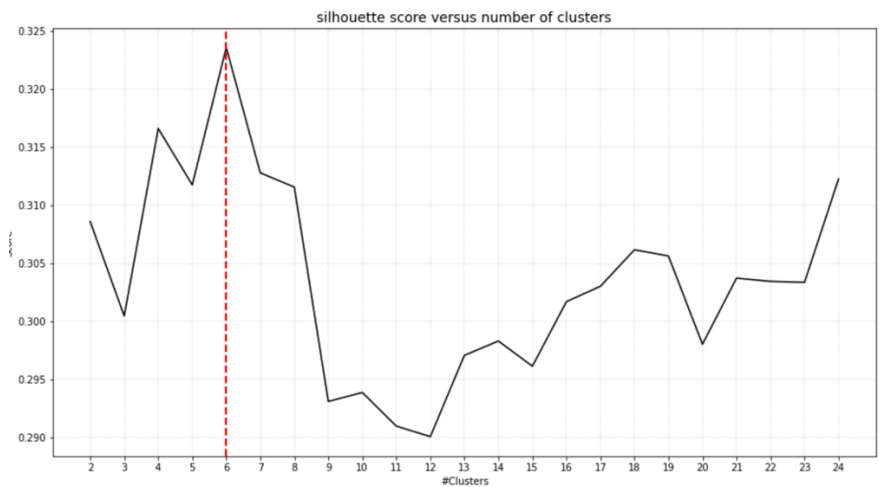

**Figure A.1. Silhouette Metric Reported for Different Number of Clusters**

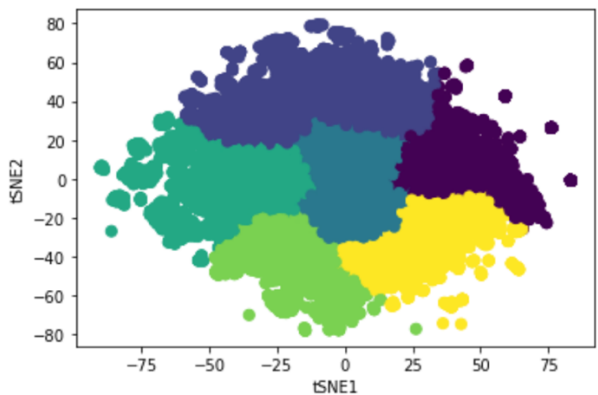

**Figure A.2. Clusters of Video Descriptions Obtained Using BIRCH**

**Reference:**
Van Der Maaten, L., and Hinton, G. 2008. "Visualizing Data Using T-SNE," *Journal of Machine Learning Research*.